\documentclass[12pt]{article} \usepackage{graphicx} \usepackage{subfigure} \usepackage{amsmath}
\usepackage{a4,color} \usepackage{jneurosci} \def\b#1{{\bf #1}} \newcommand{\bn}{{\bf n}}
\newcommand{\br}{{\bf r}} \newcommand{\di}{{\Delta I}}

\definecolor{darkgreen}{rgb}{0.2, 0.6, 0.2} \definecolor{gray}{rgb}{0.5, 0.5, 0.5}
\definecolor{orange}{rgb}{0.90, 0.50, 0.00} \definecolor{purple}{rgb}{0.60, 0.00, 0.50}

\begin{document} \begin{center}

{\bf \Large Role of correlations in population coding}

\vspace*{0.5cm}

{\large Peter E. Latham  \\} {\small Gatsby Computational Neuroscience Unit, UCL, UK}

\vspace*{0.2cm}

{\large Yasser Roudi \\} {\small Kavli Institute for Systems Neuroscience, NTNU, Trondheim, Norway\\
NORDITA, Roslagstullbacken 23, 10691 Stockholm, Sweden} \end{center}

\section{Introduction}

Correlations among spikes, both on the same neuron and across neurons, are ubiquitous in the brain.
For example cross-correlograms can have large peaks, at least in the periphery
\cite{rodieck67,Mastronarde83I,Mastronarde83II,Nirenberg01,Dan98}, and smaller -- but still
non-negligible -- ones in cortex (see \citenoparens{Cohen11} for a review), and auto-correlograms
almost always exhibit non-trivial temporal structure at a range of timescales
\cite{kim90,bair01,Deger2011}. Although this has been known for over forty years, it's still not
clear what role these correlations play in the brain -- and, indeed, whether they play any role at
all. The goal of this chapter is to shed light on this issue.

\begin{figure}
\renewcommand{\baselinestretch}{1} 
\begin{center}
\includegraphics[]{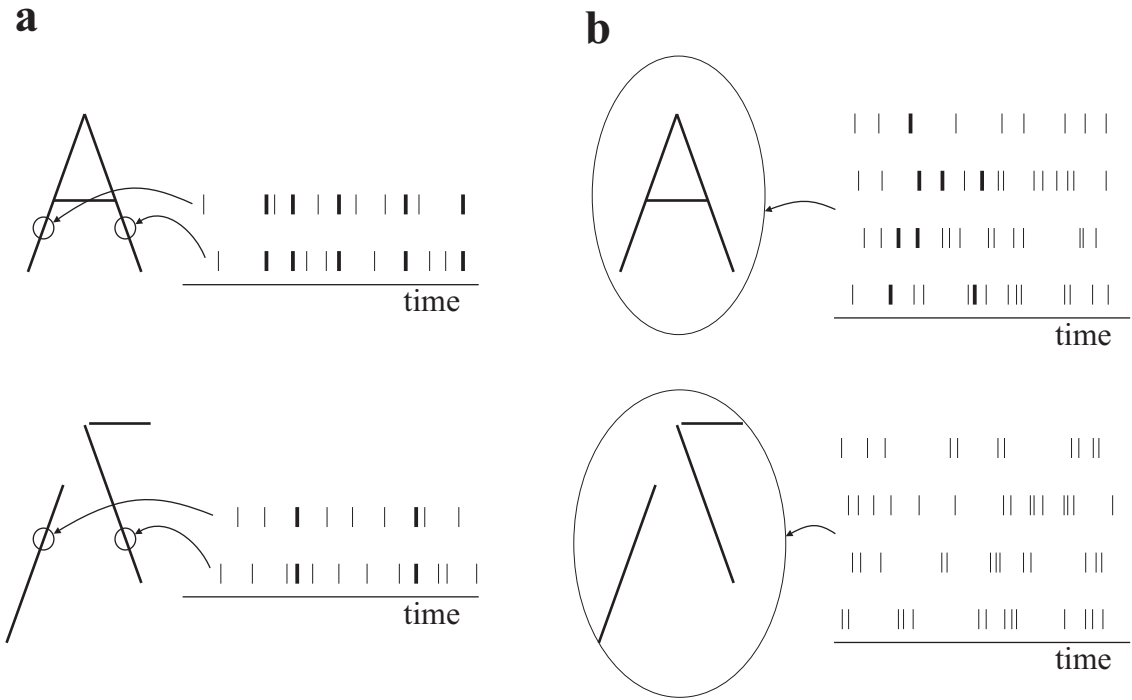} 
\caption{ Illustrative examples of synchrony and spike pattern
codes. \b a. Synchrony code. The circles in the top and bottom panels represent the receptive fields
of two neurons; the vertical bars represent spike times. In the top panel, the neurons are activated
by lines associated with a well known symbol, the letter A; to signal this, the neurons emit a large
number of synchronous spikes (thick bars). In the bottom panel, the neurons are activated by a
mainly random set of lines, and so the number of synchronous spikes (again shown as thick bars) is
at chance. 
(in this case the position, orientation and length of the set of lines), and there is an additional
spike pattern code for well known symbols. In the top panel, the image is, again, an A; the
associated spike pattern code for the A is indicated by thick bars. In the bottom panel the lines
are mainly random, and so there is no additional spike pattern code. } 
\label{meaning} 
\end{center}
\end{figure}

If synchronous spikes, or other temporal structures, are to play a role in the brain, they must
convey something of interest -- either about the outside world or about some internal state. One
example of this comes from the so-called ``binding hypothesis''
\cite{milner74,vonderMalsburg81,gray99} in which the rate of synchronous spikes across two neurons
tells us whether two objects should be bound together (Fig.~9.\ref{meaning}a). Alternatively,
arbitrary patterns of spikes, rather than just synchronous ones, could be used to signal information
about the outside world \cite{Staude10} (Fig.~9.\ref{meaning}b). In both cases, the spike patterns
act as an {\em extra} channel of information; information is also carried by overall firing rate.

A key feature of these codes is that the degree of synchrony or the rate at which patterns occur is
stimulus modulated, which automatically implies that the degree of correlations in the spike trains
is stimulus modulated. For this reason, we refer to these as correlation-based neural codes. (Here
when we refer to correlations we mean correlations conditioned on stimulus -- so-called noise
correlations, a point that is covered in detail in Chapter XI. In addition, we use the standard
neuroscience convention, which is that correlations refer to correlations at all orders, not just
covariance, as is common in the statistics literature.). To demonstrate that such correlation-based
neural codes exist, then, one merely needs to look for stimulus-modulated correlations. This,
however, is harder than it sounds, primarily because if synchronous spikes -- or spike patterns --
are an important component of the neural code (at least important from an information-theory
perspective), they need to carry information not carried in firing rate. If, for instance, whenever
the rate of synchronous spikes increases, so do firing rates, the brain could look at firing rate --
which is, typically, far easier to estimate -- and ignore synchronous spikes altogether. Of course,
it would not {\em have} to do this; it could instead look at synchronous spikes, and ignore firing
rate. However, demonstrating that experimentally is highly nontrivial. If, on the other hand,
synchronous spikes carry extra information, then if the brain wants that information, it {\em has
to} pay attention to correlations.

In this chapter we discuss quantitative methods for determining what role correlations play. By way
of background, we start with a general discussion of the neural coding problem -- which is to
determine what aspects of spike trains carry information. Here we take the point of the brain, so
``important'' means ``modify the brain's view of what is going on in the outside world.'' We then
introduce a measure of the importance of correlations to the brain, and discuss what has been found
using this measure, what it does (and doesn't) mean, and how it can be estimated for large
population. We end with a brief discussion of future directions.

\section{The neural coding problem}

The brain receives a steady stream of sensory information from the external world, a stream that is
translated into spike trains by peripheral sensors (e.g., hair cells in the auditory system and
photoreceptors in the retina). The job of the sensory system is to make sense of those spike trains;
that is, use them to construct, either explicitly or implicitly, a representation of the outside
world. In the neural coding field, we generally use $s$ to denote sensory stimuli (e.g. sounds or
visual scenes) and $\br$ to denote neural activity. The transformation from sensory information to
spike trains, is, then, a transformation from $s$ to $\br$, and the job of the brain is to invert
that transformation, and construct a mapping from $\br$ to $s$.

An important feature of this transformation that it is stochastic: if one were to record from, say,
retinal ganglion cells (the output cells of the retina) while showing exactly the same stimulus over
and over, one would record a different pattern of spike trains each time the stimulus was shown.
This probabilistic transformation is denoted $p(\br|s)$, which we refer to as the conditional
response distribution. It is not, though, the quantity of fundamental interest to the brain; what
the brain really needs is $p(s|\br)$, the probability distribution of stimuli given responses, known
as the posterior distribution. That quantity is related to $p(\br|s)$ via Bayes' theorem,

\begin{equation}
\label{bayes} p(s|\br) = \frac{p(\br|s) p(s)}{p(\br)} 
\end{equation}
\noindent
where $p(s)$ is the prior distribution over stimuli and $p(\br)$ is the total response distribution,
given by

\begin{equation} 
\label{p_r} p(\br) = \sum_s p(\br|s) p(s) \, . 
\end{equation}
\noindent
The denominator in Eq.~\eqref{bayes}, $p(\br)$, ensures that $p(s|\br)$ is properly normalized,
meaning $\sum_s p(s) = 1$. (If the stimulus is continuous, sums over $s$ are replaced by integrals.)

An important feature of Eq.~\eqref{bayes} is that the response, $\br$, carries information about the
full {\em distribution} of stimuli, not just about one single stimulus. Thus, when we say that the
problem faced by the sensory system is to make sense of incoming spike trains, what we mean is that
the problem faced by the brain is to compute $p(s|\br)$, or at least compute an approximation to it.
Of course, there is no guarantee that the brain really does this. Instead, it could simply associate
a single stimulus with each neural response. This is an important possibility to consider, because
computing full probability distributions if far harder than estimating single values. However, it
seems to be an unlikely possibility: even for something as simple as crossing the street when there
is an oncoming car, it is necessary to estimate when the car will reach us {\em and} attach error
bars to that estimate; without error bars it would be impossible to make a good decision. And,
indeed, there is mounting evidence that the brain does take into account uncertainty when making
decisions \cite{jacobs99,ernst02,kording04,chater06,ma11}; uncertainty that can only come from
$p(s|\br)$. Here, then, we consider full posteriors.

So far this is all very straightforward. However, although Eq.~\eqref{bayes} is highly compact, it
is not easy to work with, for two reasons. First, the set of stimuli is infinite, and it has a
structure that is very hard to capture mathematically. This makes it effectively impossible to
determine the probability of every stimulus (imagine, for example, trying to determine the
probability of every possible image). Consequently, the prior, $p(s)$, cannot be known, let alone
written down. Second, the response, $\b r$, consists of a set of spike {\em times}, and so lives in
an infinite dimensional space. Thus, the same pattern of activity never occurs twice, making it
impossible to estimate $p(\b r|s)$ from data if we stick to a pure spike time representation.

The first problem we ignore altogether (a common, although not universal, strategy in the neural
coding field): rather than considering realistic stimuli, we consider a relatively small set of
discrete stimuli, and simply assign a probability to each of them. For example, when investigating
the visual system, we might show images consisting of a set of oriented bars at 12 different angles,
and assign each of them a probability of 1/12.

This is a huge simplification, because it means if we know $p(\br|s$), computing $p(s|\br)$ via
Bayes' theorem is straightforward. Thus, we can focus solely on $p(\br|s)$. However, it too is very
high dimensional, so estimating it is still a hard problem.
Importantly, the brain has the same hard problem: it, like us, has to learn what spike times mean --
that is, learn how to translate from responses to stimuli, via Eq.~\eqref{bayes}.
Since it never sees the same set of spike trains twice, it can't use a pure spike time
representation; if it did, every spike train would look new, and it would never learn anything.

The solution, of course, is to apply some sort of regularization, so that spike trains that are
close mean approximately the same thing. Because the brain is a mechanistic device, this happens
naturally (barring issues of chaotic dynamics, which is a topic in itself; see, for example,
\citenoparens{London2010}). We as neuroscientists would like to know what regularization the brain 
uses. This is the {\em neural coding problem}, and it has dominated the neural coding field for the 
last several decades.

Importantly, what regularization the brain uses -- that is, what neural code it uses -- has
consequences that go well beyond the neural coding field. Consider, for example, two possible neural
codes. In one, different spike trains are considered close if there are about the same number of
spikes on each neuron in any 100 ms interval (a spike count code); in the other, different spike
trains are considered close if most of the spikes occur within about 1 ms of each other (a spike
timing code).

Networks that compute with these two neural codes are highly likely to look very different.
Therefore, before building computational models of the brain, we need to understand what the neural
code is.

\section{Approximate distributions and the neural code}

It is, unfortunately, next to impossible to {\em directly} determine what kind of regularization the
brain uses -- that would almost require a complete theory of sensory processing. We can, however,
determine this indirectly if we're willing to make one assumption: if there is information in spike
trains, the brain uses it. With that assumption, we can try out different regularizations and see
which one provides the most information about the stimulus.

There are, basically, two ways to do this. The one that was popular in the 1980s and 90s
\cite{optican87,richmond90,bialek91,bialek92}, and continues to be used today
\cite{Ince10,kayser10}, is the direct approach, in which we choose a regularization; that is, we
define explicitly what it means for two spike trains to be close. For example, we could discretize
time into bins, and replace spike times by spike count on each neuron in each time bin. This is
equivalent to defining $\br$ to be a set of spike counts rather than spike times, and it means that
two spike trains are considered exactly the same if they have the same spike counts for all neurons
in all bins, even if the spikes occurred at different times. Analysis then proceeds by computing the
mutual information \cite{shannon49} between stimuli and responses versus bin size. (For a review of
information theory, especially how it is used in neuroscience, see Chapter XI). Other notions of
closeness that have been used are principal components
\cite{optican87,richmond90,eskandar92,gawne93,gawne96,wiener99} and a metric that measures distance
in terms of how much one has to move spikes and add or delete them to make two spike trains
identical \cite{victor96,Victor05}. Here, though, we consider only binned spike trains.

The other approach is to use an approximate distribution, which we denote $q(\br|s)$, in place of
$p(\br|s)$. This approach, which was popular in the 1960s and 70s \cite{marmarelis78}, fell out of
favor when information theory was introduced, but has gained a resurgence of popularity in the last
decade \cite{Truccolo05}. Note that it includes the previous approach, for which $q(\br|s) = p(\b
f(\br)|s)$ where $\b f(\br)$ is a mapping that respects the relevant notion of closeness (e.g., for
binned spike trains, $\b f(\br)$ maps spike times to spike counts). However, it gives us a much
broader class of approximate distributions, and thus much more flexibility. It does, though,
slightly change the neural coding problem: rather than asking, say, what bin size is relevant to the
brain, it asks what approximate distribution takes us closest to the true posterior.

Which approach is better? The answer is not immediately obvious.. There are two advantages to using
explicit regularizations. The first is that it is conceptually straightforward. The second is that
it is easy to determine how much information is lost by any particular regularization. That's
because the mapping from $\br$ to $\b f(\br)$ can at best preserve information, and usually leads to
information loss. Thus, if we binned spikes, we could ask how much information is lost as a function
of bin size. At some sufficiently small bin size we would find that almost no information is lost;
the size of this bin is the timescale that matters in the brain (assuming that the brain really
wants all available information).

There is, though, a downside to using an explicit regularization, or at least to binning spikes: it
requires a large amount of data. One reason is that as the bin size get smaller, it becomes harder
and harder to estimate the probability of a spike in any one bin: the number of trials required to
accurately estimate that probability is inversely proportional to the bin size (which follows
because for small enough bin size, spiking becomes Bernoulli). And if the probability can't be
estimated accurately, the information can't be computed accurately. A related reason has to do with
correlations: the number of spikes in one bin is correlated with the number of spikes in other bins.
Thus, we really need the joint probability of spiking across bins. If, for example, we had $10$
bins, at small enough bins sizes that there could be at most one spike in each of them, there are
1024 ($2^{10}$) possible responses, and we need to compute the probability of each of them -- a
daunting task. The problem becomes exponentially harder as the number of neurons increases, because
with each neuron we get 10 more bins. Even for only 10 neurons, there are 100 bins, and so about
$10^{30}$ possible spike patterns -- and, again, we need to compute the probability of each of them!
This exponential increase is the curse of dimensionality, and it's what makes population coding so
hard.

For that reason, we are typically better off using an approximate distribution rather than directly
regularizing spike trains. This gives us the freedom to choose a parametrization for which the
number of parameters does not grow exponentially with the number of neurons. Instead, typically it
can be chosen to grow linearly or quadratically, making it feasible to determine the approximate
distribution from data. Of course, this approach also has a downside. When we bin spikes and compute
the probability distribution over spike counts, we have only two sources of error: estimation error
and error associated with the information we have thrown away by using finite bin sizes. When we use
an approximate distribution, we are typically using the {\em wrong} distribution. This requires us
to be careful about how we assess the quality of the approximate distribution we use. This is the
subject of the next section.

\section{Assessing the quality of approximate distributions: $\di$}

In practice, we often (although not always) use a mix of the two approaches described in the
previous section: we bin spikes, and then, based on the resulting spike count code, we use an
approximate distribution. For simplicity, here we discretize time into only one bin, so that $\br
\rightarrow \bn \equiv (n_1, n_2, ..., n_N)$ where $n_i$ is the spike count on neuron $i$ and there
are $N$ neurons. (We could, of course, discretize time into multiple bins, but this would add
nothing conceptually; the only effect would be to turn the $n_i$ into vectors of spike counts.) The
approximation conditional response distribution is, then, given by $q(\bn|s)$ and, in a slight abuse
of notation, we define the true distribution to be $p(\bn|s)$. Note that this is only a slight
abuse: $p(\bn|s)$ is the true distribution of spike counts; it just doesn't tell us the true
distribution over spike times.

Associated with the approximate conditional response distribution, $q(\bn|s)$, is an approximate
posterior; in analogy to Bayes' theorem, Eq.~\eqref{bayes}, it is given by

\begin{equation} 
q(s|\bn) = \frac{q(\bn|s) p(s)}{q(\bn)} \label{bayes_approx}
\end{equation}

\noindent where $q(\bn)$ is the approximate total response distribution, $q(\bn) = \sum_s q(\bn|s)
p(s)$. Note that we are using the correct prior. It too could be approximated, but we do not discuss
that here.

This leaves us with two questions: How do we determine how close $q(\bn|s)$ is to the true posterior
distribution, $p(\bn|s)$? And what approximate conditional response distribution do we use? There is
no one answer to the first question, as there are many ways to compare distributions, and the
correct one should depend on the goal of the organism under study. One approach, proposed by Amari
and colleagues, is based on the Fisher information available to an approximate decoder
\cite{Wu00,wu01}. That measure, however, can not be used with discrete stimuli, so here we use a
somewhat generalized version of their measure. It is based on what is probably the most natural
measure of distance between probability distributions, the Kullback-Leibler distance, denoted
$D_{KL}\big(p(s|\bn) || q(s|\bn) \big)$. This quantity (which is not a true distance
\cite{Kullback51}) is given by

\begin{equation}
\label{dkl_def} D_{KL}\big(p(s|\bn) || q(s|\bn) \big) = \sum_s p(s|\bn) \log
\frac{p(s|\bn)}{q(s|\bn)} \, . 
\end{equation}

\noindent This is the distance for a particular response; to get a response independent measure, we
average over all responses weighted by their probability of occurring. The resulting quantity,
denoted $\di$, is given by \cite{Nirenberg01,Nirenberg03,Latham05}

\begin{equation} 
\di = \sum_\bn p(\bn) D_{KL} \big( p(s|\bn) || q(s|\bn) \big) \, . \label{di}
\end{equation}

\noindent Note that because $\di$ is based on the Kullback-Leibler distance, it is zero only if
$q(s|\bn) = p(s|\bn)$ for all stimuli; if $q(s|\bn) \ne p(s|\bn)$ for even one stimulus, it is
positive. Throughout most of this chapter we use $\di$ as our measure of the quality of an
approximate distribution. Below we discuss in more depth its meaning, its limitations, and, briefly,
other possible measures.

The second question, ``what approximate distribution do we use?'', doesn't have an easy answer, in
large part because there are essentially no restrictions on this distribution. To choose a sensible
approximation, we need a better handle on what question we're interested in. For that we take a
close look at population coding.

\section{Population coding}

A potentially interesting, and potentially powerful, feature of population coding is the possibility
of nontrivial structure in spike trains, as discussed in the introduction in the context of the the
binding hypothesis and spike pattern codes (see in particular Fig.~9.\ref{meaning}). If  such
nontrivial structure does exist, there are several far-reaching consequences. Probably the most
important -- and often overlooked -- is that the brain must have the machinery to generate
synchronous spikes or precise patterns of activity that carry information about the stimulus.
Consequently, whether or not such structures exist is an extremely important question, since it
affects, in a very fundamental way, how we think about how the brain carries out computations.

A second, also important, consequence is that if there really is nontrivial structure in the spike
trains, it means that neurons act together to represent the world. Uncovering that representation
requires, at the very least, paired recordings -- and in the case of spike pattern codes,
simultaneous recordings from a potentially large number of neurons. Thus, computational issues
aside, from a purely pragmatic point of view it is important to know whether they exist.

The fact that nontrivial structure (or at least nontrivial structure as we have defined it here) can
be seen only when neurons are recorded simultaneously suggests a natural approximate distribution:
one in which simultaneously recorded neurons are replaced with neurons recorded on separate trials.
This is equivalent to assuming independence, for which the approximate conditional response
distribution, which we denote $q_{ind}(\bn|s)$, is given by

\begin{equation} 
\label{q_ind} q_{ind}(\bn|s) = \prod_i p(n_i|s) 
\end{equation}

\noindent where $p(n_i|s)$ is the single neuron conditional response distribution. Here the
parametrization is the single neuron conditional response distribution under a spike count
assumption. This has a very convenient feature: it removes the curse of dimensionality. That's
because if there are $k$ possible responses for each neuron ($n_i$ can take on $k$ different values)
and $N$ neurons, then for each stimulus, $s$, we need only $N(k-1)$ numbers to fully characterize
$q_{ind}(\bn|s)$ (we need $N(k-1)$ numbers rather than $Nk$ because $p(n_i|s)$ is a normalized
probability distribution). For even moderate $N$, this is many orders of magnitude smaller than the
$k^N$ numbers (more accurately, $k^N-1$, again because of the normalization) we typically need to
characterize the full distribution, $p(\bn|s)$.

By using the independent distribution as the approximate one, it seems that we are getting to the
heart of the question ``are correlations important?''. Indeed, suppose $q_{ind}(s|\bn) = p(s|\bn)$
for all stimuli (recall that $q_{ind}(s|\bn)$ is given by Eq.~\eqref{bayes_approx}). In that case,
downstream areas in the brain could both decode responses perfectly and perform computations
optimally without knowing anything about correlations. In the opposite case, $q_{ind}(s|\bn) \ne
p(s|\bn)$ for at least one stimulus, the brain would have to know about correlations. We are
tempted, then, to make the statement ``correlations are unimportant if $q_{ind}(s|\bn) = p(s|\bn)$,
and important otherwise.'' Alternatively, because $\di$ is zero if and only if $q_{ind}(s|\bn) =
p(s|\bn)$ for all stimuli (see Eq.~\eqref{di}), this statement is equivalent to ``correlations are
unimportant if and only if $\di = 0$.''

Indeed, from a purely information theory and optimal computing point of view, this is correct.
However, from the point of view of the brain, the situation is more nuanced. In Sec.~\ref{take2} we
expand on this point. First, though, we need a better understanding of some of the mathematical
properties of $\di$, and we also need to consider alternative approaches. In the next several
sections, then, we provide an information-theoretic interpretation of $\di$, provide examples in
which the responses are highly correlated and $\di = 0$, take a look at the experimental data on
$\di$, consider two alternative approximate distributions, and look at another measure of the role
of correlations.

\section{An information-theoretic perspective on $\di$}

If we did an experiment and found that $\di = 0$, the result would be easy to interpret: the
posterior distribution over stimuli computed from the independent conditional distribution is
exactly the same as that computed from the true conditional distribution. However, experimentally we
almost never find that $\di=0$; besides the fact that correlations almost always play some role,
even if $\di$ really were zero, when computed from finite data it typically becomes positive. So how
do we interpret positive values of $\di$? Our favorite interpretation is that it is the penalty one
pays, in yes/no question, in guessing the stimulus (an interpretation that is independent of whether
or not $q(\bn|s)$ is the independent distribution, $q_{ind}(\bn|s)$, or some other one). This
interpretation is discussed in some detail in \citenoparens{Nirenberg03}; here we review it briefly.

Suppose that rather than computing the posterior distribution over stimuli, we guessed the stimuli
using yes/no questions. An example of an allowed question, in the case of four stimuli, is ``is it
stimulus 1, 3 or 4?''. Once that question is answered, we get to ask another one, and the process
continues until we know exactly what the stimulus is. The optimal question asking strategy is to
divide the total probability of the stimulus in half with each question. This is a generalization of
the approach to answering the question ``We're thinking of a number between 1 and 1024; what is
it?''. Assuming a uniform prior, the first question would be something like ``is it between 1 and
512?'', and each subsequent questions would reduce the number of possibilities by a factor of two.
For a non-uniform prior, however, the strategy is different. If, for instance, you knew that we
always chose a number between 1 and 512, the first question would be something like ``is it between
1 and 128?''.

While this seems like a rather artificial approach to neural coding, it turns out that it can be
directly related to mutual information. In fact, the difference in the average number of yes/no
questions it takes to guess the stimulus before and after receiving a response {\em is} the mutual
information (a statement that is largely correct, but requires some caveats associated with batch
coding \cite{CoverandThomas}). To guess the stimulus in the minimum number of questions, one has 
to know the true posterior; if an approximate posterior is used, it will take more questions. This
was obvious in the previous example: if the question-asker had not known that we always choose 
a number between 1 and 512, she would have taken one extra question. That this is true in general 
is slightly less obvious, but is not hard to show; see, for example, \cite{CoverandThomas}. It also 
suggests a natural metric against which $\di$ should be measured: the true mutual information, 
denoted $I$, which is the reduction in yes/no questions one gets by observing the 
stimulus \cite{CoverandThomas,Nirenberg03}.

Note that while $\di$ is a yes/no question cost, and yes/no questions look a lot like bits, it's not
a true information loss. In fact, $\di$ can exceed the information \cite{schneidman03a}, something
that happens when the approximate distribution is, on average, worse than the prior distribution (at
least as measured by yes/no questions). However, as shown in \cite{Latham05}, it is an {\em upper
bound} on information loss. Consequently, small values of $\di$ are meaningful.

Although $\di$ has a relatively intuitive interpretation, is it really necessary to bother with this
quantity? Why not simply ask how correlated responses are? The answer is that just because
correlations exist doesn't mean they are important for determining the posterior distribution over
stimuli. In other words, the question ``is $q(s|\bn)$ close to $p(s|\bn)$?'' -- which we are asking
here, and which $\di$ answers -- is very different from the question ``is $q(\bn|s)$ close to
$p(\bn|s)$?''.

\begin{figure} 
\renewcommand{\baselinestretch}{1} 
\begin{center} \includegraphics[height=11cm,width=12cm]{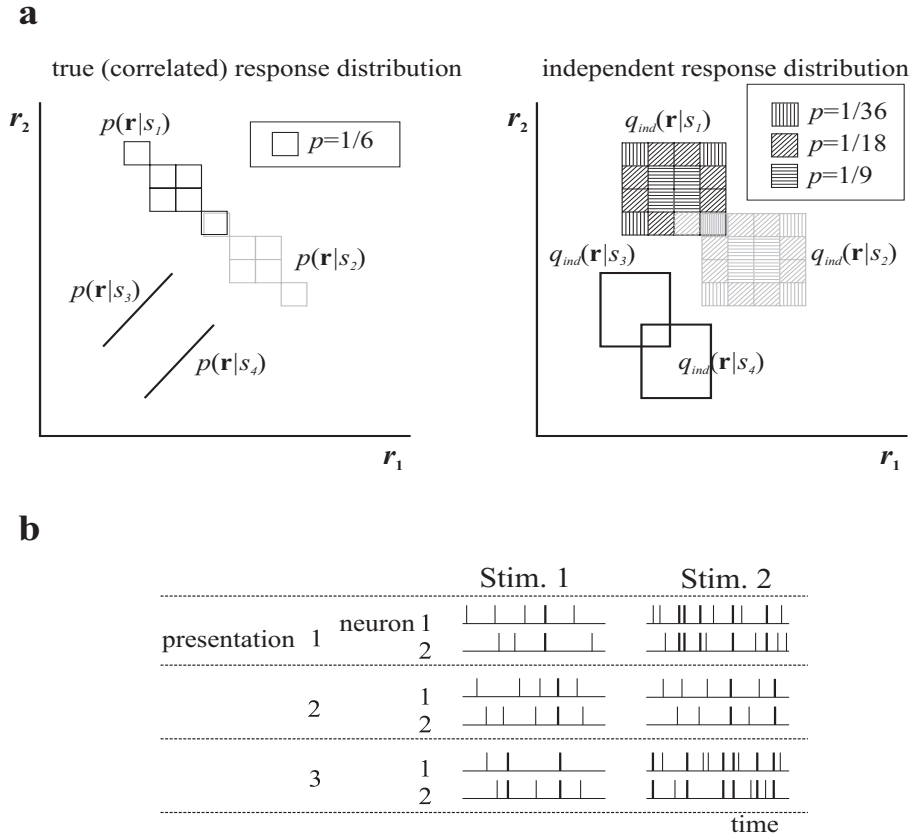}
\caption{ \small $\di$ can be zero when correlations are strong: two hypothetical examples. \b a.
Continuous distributions for four stimuli. The left panel shows the true conditional response
distribution. For stimuli 1 and 2 the probability is uniform within each square, and the probability
of landing within any one of the squares is 1/6, as indicated in the legend. The lower right square
associated with stimulus 1 perfectly overlaps the upper left square associated with stimulus 2; they
are offset slightly for clarity. For stimuli 3 and 4 the responses are perfectly correlated: $r_1$
perfectly predicts $r_2$ and vice-versa. The right panel shows the independent conditional response
distribution. For stimuli 1 and 2 the squares now have different probabilities, as indicated in the
legend, and none of them are equal to the true probabilities. However, in the overlap region the
ratio of the probabilities remains the same and so, via Bayes' theorem, the responses would be
decoded perfectly. For example, assuming uniform priors, if the response fell in the overlap region,
under both the true and independent conditional response distributions the probability assigned to
stimuli 1 and 2 would be 1/2. For stimuli 3 and 4 the conditional response distributions are now
uniform within the rectangles. Note that if a response fell in the overlap region it would not be
clear which stimulus caused it. However, under the true distribution the responses never fall in the
overlap region, so even if one used the independent conditional response distribution to decode, one
would decode perfectly all true responses. \b b. Spiking response and two stimuli. Synchronous
spikes are indicated by thick bars. Stimulus 2 produces a higher firing rate than stimulus 1, but it
also produces more synchronous spikes. As is not hard to show, if the probability of synchronous
spikes is a function of the firing rates only, and does not depend on the stimulus, then knowledge
of synchronous spikes adds no information, and $\di$ is zero. Adapted from Nirenberg and Latham,
2003. }
\label{di=0} 
\end{center} 
\end{figure}

A general mathematical prescription for when $q(s|\bn)$ and $p(s|\bn)$ can be equal even though
$q(\bn|s)$ and $p(\bn|s)$ are different was provided by \citenoparens{amari06}. Intuitively, though,
it's easy to see why -- and when -- this can happen. What matters is that in regions where the same
response can lead to different stimuli, the {\em relative} response probabilities associated with
different stimuli are the same under the approximate and true distributions. This is illustrated in
Fig.~9.\ref{di=0}a for a two dimensional response distribution when the approximate distribution is
the independent one, $q_{ind}(\bn|s)$, and in Fig.~9.\ref{di=0}b for spike trains, again when the
approximate distribution is the independent one. In both cases $q_{ind}(\bn|s)$ and $p(\bn|s)$ are
very different even though $q_{ind}(s|\bn)$ and $p(s|\bn)$ are identical.

An additional example in which correlations exist but do not affect the posterior comes from the
linear probabilistic population coding framework \cite{pouget03,ma06}. In this framework,
conditional response distributions have the form

\begin{equation}
\label{ppc} p({\bf n}|s) = \phi({\bf n}) \exp\big({\bf h}(s) \cdot {\bf n} +
\psi(s)\big)
\end{equation}

\noindent where $\phi(\bn)$ is an arbitrary function of the responses and $\b h(s)$ and $\psi(s)$
are arbitrary functions of the stimulus. Here the correlations, which are contained in $\phi(\bn)$,
could be very complicated, but we don't have to know them to compute the posterior. As is easy to
show,

\begin{equation} 
p(s|{\bf n}) = \frac{\exp \big({\bf h}(s) \cdot {\bf n} + \psi(s) \big) p(s)}
{\sum_{s'} \exp \big({\bf h}(s') \cdot {\bf n} + \psi(s') \big) p(s')} \, , 
\end{equation}

\noindent which is independent of $\phi(\bn)$, and thus does not require knowledge of the
correlations. That we don't need to know the correlational structure to determine $p(s|\bn)$ is not
so surprising in this case, since the correlations are stimulus independent. Nevertheless, if this
-- or a close approximation to it -- really is the distribution used by the brain, then we don't
have to worry about correlations at all, even when they are quite strong (i.e., when $\phi(\bn)$
yields a highly correlated distribution).

\section{What experiments tell us}

What value of $\Delta I$ does one find in experimental data? The first study to address this
question was performed by Yang Dan and colleagues, in the cat lateral geniculate nucleus
\cite{Dan98}. They found that, for some pairs of neurons, using the independent distribution
resulted in a 40\% information loss. This was not, however, the general trend. Studies in rat barrel
cortex \cite{Petersen01}, mouse retina \cite{Nirenberg01}, V1 \cite{Golledge03}, motor cortex
\cite{Averbeck06} and supplementary motor  \cite{Averbeck03} area all showed that $\di$ was of the
order of $10\%$ of the total information. In the one case in which the information loss was measured
(as opposed to $\di$, which is an upper bound) from experimental data, it was very close to $\di$
\cite{oizumi09,oizumi10}.

There is, though, one other study in which large $\di$ was found \cite{Ince10}. In that study, the
authors found that for pairs or triplet of neurons in the barrel cortex, $\di$ was small. However,
by increasing the population size to $8$, $\di$ became significant, on the order of $40\%$ of the
total information. As the authors note, this could mean that the the neurons receiving input from
the barrel cortex must know about the correlations between barrel neurons.

\section{What to do when investigating large populations}

One problem with $\di$ is that it can't be computed from data for more than a handful of neurons.
That's because it depends on the true conditional response distribution (see Eq.~\eqref{di}), which
we typically don't know, and, indeed, which the curse of dimensionality tells us we can't know. What
we can do, though, is consider families of parametrized distributions, and ask whether the posterior
distribution converges as the family becomes more complex; we just can't ask if it converges to the
true posterior. Two such families commonly used in neuroscience are generalized linear models, or
GLMs \cite{Truccolo05,Pillow08} and maximum entropy models
\cite{Jaynes57I,Jaynes57II,schneidman03b,Schneidman06,Shlens06}. Here we describe them briefly, and
summarize what we have learned from them.

Although GLMs and maximum entropy models share some similarities, there are two major differences.
The first has to do with bin size. While GLMs are often used with finite bin sizes, this is not
necessary, and in fact one of the strengths of these models is that they make sense in the
continuous time limit \cite{gerwinn10}. For maximum entropy models, on the other hand, results
depend critically on bin size \cite{Roudi09}. The second difference has to do with how past and
current spikes influence the probability of spiking: in GLMs, past spikes have a strong influence
and current spikes have none; in maximum entropy models (at least the versions usually used in
neuroscience) it is just the opposite.

To make this explicit, we start by writing down the conditional response distribution for GLMs. In
these models, the probability of spiking is independent conditioned on the stimulus and previous
spikes. Discretizing time (to better make contact with maximum entropy models) and suppressing the
dependence on previous spikes (for ease of notation), we have, therefore,

\begin{equation} 
\label{glm0} q({\bf n}(t)|s) = \prod_i q(n_i(t)|s) \, .
\end{equation}

\noindent The individual distributions, $q(n_i(t)|s)$, are given by

\begin{equation} 
\label{glm} q(n_i(t)|s) = \frac{1}{Z} \exp \left[ K_i[s] n_i(t) + \sum_{t' < t}
h_i(t')n_i(t-t') + \sum_{t' < t, j \ne i} J_{ij}(t') n_j(t-t') \right] \, . 
\end{equation}

\noindent where $Z$ ensures that $q(n_i(t)|s)$ is properly normalized. Here the notation $t' < t$
indicates that the sum contains only previous bins (and, recall, time is a discrete variable; thus
the sum). This sum, of course, extends only a finite time into the past. The dependence on the
stimulus is essentially arbitrary, but it is usually taken to be a temporal linear convolution, or,
if the stimulus is spatially varying, a spatio-temporal linear convolution (the brackets around $s$
indicate that there is a dependence on stimulus history). The parameters of the GLM are $h_i(t')$
and $J_{ij}(t')$, and any parameters associated with $K_i[s]$. In addition, the nonlinearity does
not have to be exponential, but if a different nonlinearity is chosen, extra parameters are
(typically) needed to characterize it. And finally, Eq.~\eqref{glm} is technically correct only in
the limit of infinitesimally small bin size.

Maximum entropy models are a broad class of models in which one constructs the distribution that has
maximum entropy, subject to constraints. Here we consider second order maximum entropy models, since
those are the ones most commonly used in neuroscience
\cite{Schneidman06,Shlens06,tkacik06,tkacik07,tang08,yu08,shlens09,Roudi09b,Roudi09c,granot10,Ohiorhenuan10,ganmor11b,ganmor11a}. 
For these models the constraints are on the first and second
moments of the probability of spiking in a bin. When those moments can depend on stimulus, the most
common form for the model is

\begin{equation}
\label{ising} q({\bf n}(t)|s) = \frac{1}{Z(t)} \exp\left[ \sum_i h_i[s] n_i(t) +
\sum_{j \ne i} J_{ij}[s] n_i(t) n_j(t) \right] 
\end{equation} 

\noindent where $Z(t)$ ensures that
$q(\bn|s)$ is properly normalized, and, as above, the fact that the stimulus appears in square
brackets in $h_i[s]$ and $J_{ij}[t]$ indicates that these terms depend on stimulus history. Equation
\eqref{ising} has the form of the Ising model \cite{ising25}; because it also has stimulus
dependence, we call it the stimulus-dependent Ising model.

For both GLMs and Ising models, correlations across neurons are contained in the coupling terms, the
$J_{ij}$. Thus, one can assess the role of correlations by asking about the quality of the model
with and without that term. This is, of course, difficult to do exactly, but one can take an
approximate approach. Perhaps the simplest is to build a decoder based on the approximate
distribution, $q(\b n|s)$, with and without the coupling terms, and either compute its variance
numerically or, for discrete stimuli, estimate the probability of making a correct classification.
Correlations can then be assessed by comparing the decoder under the two conditions.

So what has been found using these two models? In the case of GLMs, Pillow and colleagues fit the
model to retinal ganglion cells from macaque monkeys, and estimated the signal-to-noise ratio with
and without the coupling terms \cite{Pillow08}. A smaller signal to noise ratio essentially means a
lower variance decoder. What they found was that, for populations of 27 neurons, the log of the
signal to noise was about 20\% lower when the coupling terms were excluded from the model (that is,
when correlations were ignored). Whether or not one considers this a large or small information loss
is a matter of taste; but we feel that it is small -- after all, as pointed out by Pillow et al.,
it's possible that the ratio of the information with and without the coupling terms could have
scaled linearly with the number of neurons. Had this been the case, the information loss would have
been on the order of 97\%, not 20\%. If the information loss stays at 20\% for even larger
populations, then it may be the case that correlations -- or at least pairwise correlations -- don't
have much affect on the posterior distribution over stimuli.

For Ising models, most studies have not considered any stimulus dependence
\cite{Schneidman06,Shlens06,tkacik07,tang08,yu08,shlens09,ganmor11b}. However, there are four that
have \cite{tkacik07,granot10,Ohiorhenuan10,ganmor11a}. The first of these assumed that the $h_i$,
but not the $J_{ij}$, depended on the stimulus \cite{tkacik07}. For populations of 10 retinal
ganglion cells, taking into account the correlations resulted in a modest improvement -- the
correlated model ($J_{ij} \ne 0$) did about 10\% better predicting the absence of firing compared to
the independent model ($J_{ij} = 0$). The second did not directly examine decoding, but they did
find that the stimulus dependent Ising model did a much better job predicting stimuli than the
stimulus dependent model \cite{granot10}. This was, though, the only model that allowed the
couplings, the $J_{ij}$, to depend on stimulus. Interestingly, the dependence was very weak; the
implications of that finding are not yet fully understood. In the third model, the authors did not
investigate the effect of the coupling terms on the posterior distribution over stimuli
\cite{Ohiorhenuan10}. They did compute information with and without coupling, but, as discussed in
the next section (and elsewhere \cite{Latham05}), it is not clear what this tells us about the role
of correlations. The fourth study was probably the most interesting. Here the authors considered a
maximum entropy model that went beyond second order \cite{ganmor11a}. When they used that model to
decode novel stimuli from $\sim$100 retinal ganglion cells, they could decode them about three times
faster than when they used the independent model. Thus, it seems that in this case correlations are
clearly important.

\section{Other measures of the role of correlations}

So far we have asked how correlations affect one's ability to compute the posterior distribution
over stimuli, $p(s|\br)$. However, one may ask a different question: what is the effect of
correlations on the ability of a population to encode information? There are several reasons for
asking this question. One is to gain intuition about whether correlations are ``good'' or ``bad'' --
that is, whether they increase or decrease information. The other, related, reason is that the brain
might use strategies to modify the correlations, and so measuring correlations in several conditions
(e.g., with and without attention) may give us insight into computational strategies used by the
brain.

\begin{figure} 
\renewcommand{\baselinestretch}{1} 
\begin{center}
\includegraphics[]{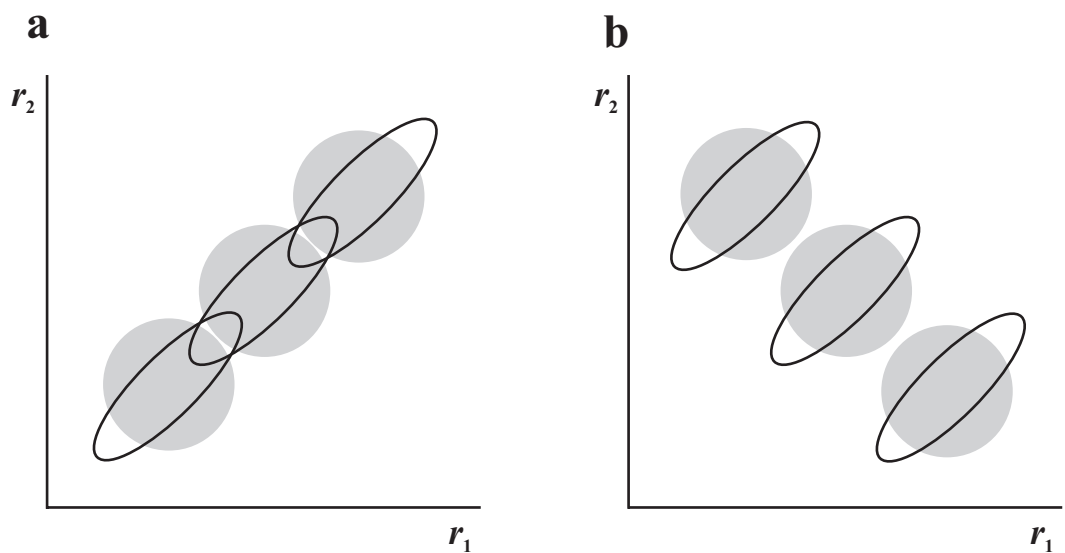}
\caption{ \small Correlations can either increase or decrease
information. In both figures, the responses are taken to be continuous and the conditional response
distributions are Gaussian. The ellipses indicate contours containing $90\%$ of the probability for
the true responses; the circles contain 90\% of the probability for the independent responses.
\small \b a. Correlations make the responses harder to decode relative to the independent responses,
so they decrease information. \b b. Correlations make the responses easier to decode relative to the
independent responses, so they decrease information. } 
\label{shuffled} 
\end{center}
\end{figure}

Theoretically, this question was addressed over a decade ago by Abbott and Dayan \cite{Abbott99},
who found that correlations can either increase or decrease information in a population code. The
reason is rather easy to see, and can be illustrated with only two neurons using a spike count code
(in which, for ease of exposition, we pretend that spike count is a continuous variable). Suppose
that conditioned on stimulus, the spike counts of the two neurons are positively correlated. In that
case, the conditional response distributions form ellipse-like shapes that are neither purely
vertical or purely horizontal, as illustrated in Figs.~9.\ref{shuffled}a and b. The amount of
information in the population depends on how easy the responses are to decode, and thus on how much
overlap there is between the responses associated with each stimuli. If the mean responses for each
of the stimuli are aligned with the long axis of the ellipses, information will be low (correlations
decrease information; Fig.~9.\ref{shuffled}a), and if the mean responses are aligned with the short
axis, information will be high (correlations increase information; Fig.~9.\ref{shuffled}b).

Despite that fact that correlations can either increase or decrease information, there seems to be a
feeling among the community that correlations generally decrease it \cite{Zohary94,Shadlen96}. While
not {\em necessarily} true, it would seem to be true whenever neurons have similar tuning properties
and neurons are positively correlated (so that the responses look more like Fig.~9.\ref{shuffled}a
than 9.\ref{shuffled}b). In particular, for a very common correlational structure -- neurons with
similar tuning are more correlated than neurons with dis-similar tuning curves -- increasing
correlations does decrease information \cite{Abbott99,Sompolinsky01} It is this intuition that is
behind several studies that looked at how correlations changed with attention. These studies found
that attention decreased them, and so, it was stated, information should go up
\cite{cohen09,mitchell09}. However, a recent study showed that if tuning curves do not all have the
same amplitude (as is probably the case in the brain), then, even when correlations are large and
positive for similarly tuned neurons and weak for dis-similarly tuned ones (exactly the case for
which correlations should decrease information), increasing correlations does not lead to much of a
decrease in information \cite{ecker11}. In fact, for a large enough population, correlations of this
type always increase information. So we're back where we started: correlations can either increase
or decrease information, and it can be very hard to make general statements about which one will
happen in realistic situations.

\section{Correlations, learning and computations: $\di$ take 2} \label{take2}

Suppose we did an experiment and found that the posterior distribution over stimuli under the
independence assumption was equal to the true posterior ($\di = 0$). We could, then, measure single
neuron conditional response distributions and use them to build optimal decoders -- and, by
extension, perform optimal computations. So far in this chapter we have {\em defined} this to mean
that correlations are not important. However, we should keep in mind that this is not the only
notion of important, or even the best one. Indeed, the above assertion comes with a number of
caveats.

\begin{figure} 
\renewcommand{\baselinestretch}{1}
\begin{center}
\includegraphics[]{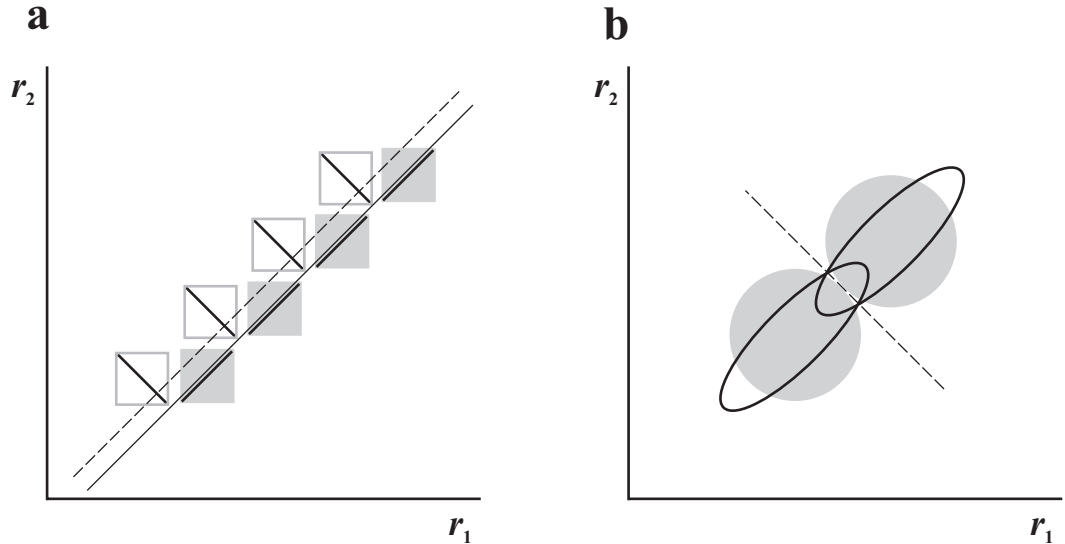} 
\caption{ \small For suboptimal decoders, $\di$ can be
misleading. \small \b a. Eight stimuli need to be divided into two classes using a linear
classifier. The responses to the stimuli in class 1 (upper left) are perfectly anti-correlated, as
indicated by the bars at $-45^\circ$; the responses to stimuli in class 2 (lower right) are
perfectly correlated, as indicated by the bars at $+45^\circ$. The independent conditional response
distributions are squares; open for class 1 and gray for class 2. The dashed line is the optimal
linear classifier under the independence assumption, for which some errors are made for class 1
stimuli. The solid line is the optimal linear classifier under the true distribution, for which
errors are never made. Thus, although $\di=0$, a suboptimal decoder does not perform perfectly. \b
b. Two stimuli need to be divided need into two classes, again with a linear classifier. The
ellipses show the true distribution; the circles show the independent one (as in
Fig.~9.\ref{shuffled}). Here $\di \ne 0$, since, for almost all responses, the posterior probability
of the stimuli is different for the true and independent assumption. However, the linear classifier
(dashed line) is the same for both the true and independent distributions. Thus, if all one cared
about was the performance of a linear classifier, the fact that $\di$ is greater than 0 would be of
little interest.}
\label{linclass} 
\end{center} 
\end{figure}

An important one is the qualifier ``optimal'' that appears above. While it's true that when $\di=0$
we can decode optimally, it is not true that we can perform approximate decoding optimally. For
example, suppose that $\di=0$ under the independence assumption, and we wanted to build a linear
classifier that divides stimuli into two classes. In Fig.~9.\ref{linclass}a we show the responses to
eight stimuli, four of which (the ones on the upper left) should be in class 1 and the other four
(the ones in the lower right) should be in class 2. Under the independence assumption, in which the
conditional response distributions are squares, the optimal linear classifier runs right between the
two classes (dashed line). Under the true conditional response distribution, this classifier is
always correct for class 2, but correct only about 75\% of the time for class 1. This is in contrast
to the optimal linear classifier (solid line), for which classification is perfect. However, to find
it, the true conditional response distribution must be known. So finding that $\di = 0$ does not
mean one can build an optimal linear classifier.

The opposite can also occur: $\di$ can be positive even when a suboptimal decoder can work perfectly
without knowledge of the correlational structure. Consider, for example Gaussian conditional
response distributions from two neurons, as shown in Fig.~9.\ref{linclass}b for two stimuli. As is
clear from this figure (or is easy to calculate), $\di$ is not zero. However, the optimal linear
classifier under the independence assumption is the same as it is for the true model.

The take home message here is that $\di$ may not be very informative about how well approximate
decoders will fare. This is especially important because the problems faced by the brain are so
complicated that it almost always (if not always) has to make some approximations. If the
approximations the brain makes is known, then finding the correct measure is easy. For example, for
a linear classifier, the correct measure would involve a comparison between the fraction correct
under the independent conditional response distribution and under the true one. However, if the
approximation the brain takes is not known, finding the correct measure is a nontrivial task.

Second, and equally, if not more, important, the brain doesn't have access to single neuron
responses. Thus, even if it wanted to construct the posterior distribution based on the independent
responses, it couldn't. What it sees are the full, correlated responses on every trial. The question
we should be asking, then, is: how do correlations affect both learning and the optimal computations
used by the network? Pouget and colleagues have addressed the second question, although for a
relatively simple problem (by the standards of the brain), cue combination \cite{ma06}. Although
they did not explicitly consider correlations, the techniques they used could be extended, with some
work, in that direction.

\section{Summary}

So what is the role of correlations? As we have seen here, experimentally one can often construct a
near-optimal posterior distribution over stimuli based only on the single neuron conditional
response distributions -- and, therefore, with no knowledge of the correlational structure
\cite{Nirenberg01,Petersen01,Averbeck03,Golledge03,Averbeck06,tkacik07,granot10}. There are, though,
exceptions \cite{Dan98,Ince10,ganmor11a}. Of these, the study by Ganmor et al.\ is especially
interesting, because it considered $\sim$100 retinal ganglion cells and found that when correlations
were taken into account, decoding speed increased by a factor of about three \cite{ganmor11a}. From
an evolutionary standpoint, such a speed increase would be highly beneficial. Whether this is a
general principle throughout the cortex remains to be seen, but in any case it makes an important
point: we should be studying large, not small, neuronal populations.

Finally, our approach -- asking whether one needs to know about correlations to accurately represent
the outside world -- isn't {\em exactly} the question we want to ask of the brain. The brain
computes with responses rather than explicitly constructing the posterior distribution, it typically
does so using approximate algorithms, and it doesn't have access to the independent conditional
response distribution. The first two are relevant because we saw that when performing approximate
computations, it may be necessary to know the true distribution even when $\di = 0$
(Fig.~9.\ref{linclass}a), and it may not be necessary to know the true distribution even when $\di
\ne 0$ (Fig.~9.\ref{linclass}b). The third is relevant because the brain has to learn what to do
based on correlated responses, and so the real question is: how do correlations affect learning?
This is a problem for which the answer is not known, at least not in general.
Thus, despite much work on the role of correlations, much is left to be done.

\section{Acknowledgments} 

We would like to thank Jonathan Pillow for valuable feedback on this
chapter.


\begin{thebibliography}{}

\bibitem[\protect\citeauthoryear{Abbott and Dayan}{1999}]{Abbott99}
Abbott LF, Dayan P (1999)
\newblock The effect of correlated variability on the accuracy of a population
  code.
\newblock {\em Neural Comput}~11:\mbox{91--101}.

\bibitem[\protect\citeauthoryear{Amari and Nakahara}{2006}]{amari06}
Amari S, Nakahara H (2006)
\newblock Correlation and independence in the neural code.
\newblock {\em Neural Comput}~18:\mbox{1259--1267}.

\bibitem[\protect\citeauthoryear{Averbeck and Lee}{2003}]{Averbeck03}
Averbeck BB, Lee D (2003)
\newblock Neural noise and movement-related codes in the macaque supplementary
  motor area.
\newblock {\em J Neurosci}~23:\mbox{7630--7641}.

\bibitem[\protect\citeauthoryear{Averbeck and Lee}{2006}]{Averbeck06}
Averbeck BB, Lee D (2006)
\newblock Effects of noise correlations on information encoding and decoding.
\newblock {\em J Neurophysiol}~95:\mbox{3633--3644}.

\bibitem[\protect\citeauthoryear{Bair \bgroup et al.\egroup }{2001}]{bair01}
Bair W, Zohary E, Newsome WT (2001)
\newblock Correlated firing in macaque visual area {MT}: time scales and
  relationship to behavior.
\newblock {\em J Neurosci}~21:\mbox{1676--1697}.

\bibitem[\protect\citeauthoryear{Bialek and Rieke}{1992}]{bialek92}
Bialek W, Rieke F (1992)
\newblock Reliability and information transmission in spiking neurons.
\newblock {\em Trends Neurosci}~15:\mbox{428--434}.

\bibitem[\protect\citeauthoryear{Bialek  \bgroup et al.\egroup
  }{1991}]{bialek91}
Bialek W, Rieke F, de~Ruyter~van Steveninck RR, Warland D (1991)
\newblock Reading a neural code.
\newblock {\em Science}~252:\mbox{1854--1857}.

\bibitem[\protect\citeauthoryear{Chater \bgroup et al.\egroup
  }{2006}]{chater06}
Chater N, Tenenbaum JB, Yuille A (2006)
\newblock Probabilistic models of cognition: where next?
\newblock {\em Trends Cogn Sci}~10:\mbox{292--293}.

\bibitem[\protect\citeauthoryear{Cohen and Kohn}{2011}]{Cohen11}
Cohen MR, Kohn A (2011)
\newblock Measuring and interpreting neuronal correlations.
\newblock {\em Nat Neurosci}~14:\mbox{811--819}.

\bibitem[\protect\citeauthoryear{Cohen and Maunsell}{2009}]{cohen09}
Cohen MR, Maunsell JH (2009)
\newblock Attention improves performance primarily by reducing interneuronal
  correlations.
\newblock {\em Nat Neurosci}~12:\mbox{1594--1600}.

\bibitem[\protect\citeauthoryear{Cover and Thomas}{1991}]{CoverandThomas}
Cover T, Thomas J (1991)
\newblock {\em Elements of Information theory.}
\newblock New York: Wiley.

\bibitem[\protect\citeauthoryear{Dan  \bgroup et al.\egroup }{1998}]{Dan98}
Dan Y, Alonso JM, Usrey WM, Reid RC (1998)
\newblock Coding of visual information by precisely correlated spikes in the
  {LGN}.
\newblock {\em Nature Neurosci}~1:\mbox{501--507}.

\bibitem[\protect\citeauthoryear{Deger  \bgroup et al.\egroup
  }{2011}]{Deger2011}
Deger M, Helias M, Boucsein C, Rotter S (2011)
\newblock Statistical properties of superimposed stationary spike trains.
\newblock {\em J Comp Neurosci}~in press.

\bibitem[\protect\citeauthoryear{Ecker  \bgroup et al.\egroup }{2011}]{ecker11}
Ecker A, Berens P, Tolias A, Bethge M (2011)
\newblock The effect of noise correlations in populations of diversely tuned
  neurons.
\newblock {\em J Neurosci}~in press.

\bibitem[\protect\citeauthoryear{Ernst and Banks}{2002}]{ernst02}
Ernst MO, Banks MS (2002)
\newblock Humans integrate visual and haptic information in a statistically
  optimal fashion.
\newblock {\em Nature}~415:\mbox{429--433}.

\bibitem[\protect\citeauthoryear{Eskandar \bgroup et al.\egroup
  }{1992}]{eskandar92}
Eskandar EN, Richmond BJ, Optican LM (1992)
\newblock Role of inferior temporal neurons in visual memory. {I}. temporal
  encoding of information about visual images, recalled images, and behavioral
  context.
\newblock {\em J Neurophysiol}~68:\mbox{1277--1295}.

\bibitem[\protect\citeauthoryear{Ganmor \bgroup et al.\egroup
  }{2011a}]{ganmor11b}
Ganmor E, Segev R, Schneidman E (2011a)
\newblock The architecture of functional interaction networks in the retina.
\newblock {\em J Neurosci}~31:\mbox{3044--3054}.

\bibitem[\protect\citeauthoryear{Ganmor \bgroup et al.\egroup
  }{2011b}]{ganmor11a}
Ganmor E, Segev R, Schneidman E (2011b)
\newblock Sparse low-order interaction network underlies a highly correlated
  and learnable neural population code.
\newblock {\em Proc Natl Acad Sci USA}~108:\mbox{9679--9684}.

\bibitem[\protect\citeauthoryear{Gawne  \bgroup et al.\egroup }{1996}]{gawne96}
Gawne TJ, Kjaer TW, Hertz JA, Richmond BJ (1996)
\newblock Adjacent visual cortical complex cells share about 20\% of their
  stimulus-related information.
\newblock {\em Cereb Cortex}~6:\mbox{482--489}.

\bibitem[\protect\citeauthoryear{Gawne and Richmond}{1993}]{gawne93}
Gawne TJ, Richmond BJ (1993)
\newblock How independent are the messages carried by adjacent inferior
  temporal cortical neurons?
\newblock {\em J Neurosci}~13:\mbox{2758--2771}.

\bibitem[\protect\citeauthoryear{Gerwinn \bgroup et al.\egroup
  }{2010}]{gerwinn10}
Gerwinn S, Macke JH, Bethge M (2010)
\newblock Bayesian inference for generalized linear models for spiking neurons.
\newblock {\em Front Comput Neurosci}~4.

\bibitem[\protect\citeauthoryear{Golledge  \bgroup et al.\egroup
  }{2003}]{Golledge03}
Golledge HD, Panzeri S, Zheng F, Pola G, Scannell W, Giannikopoulos DV, Mason
  RJ, Tovee MJ, Young MP (2003)
\newblock Correlations, feature-binding and population coding in primary visual
  cortex.
\newblock {\em Neuroreport}~14:\mbox{1045--1050}.

\bibitem[\protect\citeauthoryear{Granot-Atdegi  \bgroup et al.\egroup
  }{2010}]{granot10}
Granot-Atdegi E, Tka\v{c}ik G, Segev R, Schneidman E (2010)
\newblock A stimulus-dependent maximum entropy model of the retinal population
  neural code
\newblock In {\em Frontiers in Neuroscience}. Computational and Systems
  Neuroscience 2010.

\bibitem[\protect\citeauthoryear{Gray}{1999}]{gray99}
Gray C (1999)
\newblock The temporal correlation hypothesis of visual feature integration:
  still alive and well.
\newblock {\em Neuron}~24:\mbox{31--47, 111--125}.

\bibitem[\protect\citeauthoryear{Ince  \bgroup et al.\egroup }{2010}]{Ince10}
Ince R, Senatore R, Arabzadeh E, Montani F, E. DM, Panzeri S (2010)
\newblock Information-theoretic methods for studying population codes.
\newblock {\em Neural Networks}~23:\mbox{713--727}.

\bibitem[\protect\citeauthoryear{Ising}{1925}]{ising25}
Ising E (1925)
\newblock Beitrag zur theorie des ferromagnetismus.
\newblock {\em Z Physik}~31:\mbox{253--258}.

\bibitem[\protect\citeauthoryear{Jacobs}{1999}]{jacobs99}
Jacobs RA (1999)
\newblock Optimal integration of texture and motion cues to depth.
\newblock {\em Vision Res}~39:\mbox{3621--3629}.

\bibitem[\protect\citeauthoryear{Jaynes}{1957a}]{Jaynes57I}
Jaynes ET (1957a)
\newblock Information theory and statistical mechanics.
\newblock {\em Physical Review Series {II}}~106:\mbox{620--630}.

\bibitem[\protect\citeauthoryear{Jaynes}{1957b}]{Jaynes57II}
Jaynes ET (1957b)
\newblock Information theory and statistical mechanics {II}.
\newblock {\em Physical Review Series {II}}~108:\mbox{171--190}.

\bibitem[\protect\citeauthoryear{Kayser \bgroup et al.\egroup
  }{2010}]{kayser10}
Kayser C, Logothetis NK, Panzeri S (2010)
\newblock Millisecond encoding precision of auditory cortex neurons.
\newblock {\em Proc Natl Acad Sci USA}~107:\mbox{16976--16981}.

\bibitem[\protect\citeauthoryear{Kim \bgroup et al.\egroup }{1990}]{kim90}
Kim DO, Sirianni JG, Chang SO (1990)
\newblock Responses of {DCN-PVCN} neurons and auditory nerve fibers in
  unanesthetized decerebrate cats to {AM} and pure tones: analysis with
  autocorrelation/power-spectrum.
\newblock {\em Hear Res}~45:\mbox{95--113}.

\bibitem[\protect\citeauthoryear{K{\"o}rding and Wolpert}{2004}]{kording04}
K{\"o}rding KP, Wolpert DM (2004)
\newblock Bayesian integration in sensorimotor learning.
\newblock {\em Nature}~427:\mbox{244--247}.

\bibitem[\protect\citeauthoryear{Kullback and Leibler}{1951}]{Kullback51}
Kullback S, Leibler R (1951)
\newblock On information and sufficiency.
\newblock {\em Annals of Mathematical Statistics}~22:\mbox{79--86}.

\bibitem[\protect\citeauthoryear{Latham and Nirenberg}{2005}]{Latham05}
Latham PE, Nirenberg S (2005)
\newblock Synergy, redundancy, and independence in population codes, revisited.
\newblock {\em J Neurosci}~25:\mbox{5195--5206}.

\bibitem[\protect\citeauthoryear{London  \bgroup et al.\egroup
  }{2010}]{London2010}
London M, Roth A, Beeren L, Hausser M, Latham PE (2010)
\newblock Sensitivity to perturbations in vivo implies high noise and suggests
  rate coding in cortex.
\newblock {\em Nature}~466:\mbox{123--127}.

\bibitem[\protect\citeauthoryear{Ma  \bgroup et al.\egroup }{2006}]{ma06}
Ma WJ, Beck JM, Latham PE, Pouget A (2006)
\newblock Bayesian inference with probabilistic population codes.
\newblock {\em Nat Neurosci}~9:\mbox{1432--1438}.

\bibitem[\protect\citeauthoryear{Ma  \bgroup et al.\egroup }{2011}]{ma11}
Ma WJ, Navalpakkam V, Beck JM, Berg R, Pouget A (2011)
\newblock Behavior and neural basis of near-optimal visual search.
\newblock {\em Nat Neurosci}~14:\mbox{783--790}.

\bibitem[\protect\citeauthoryear{Marmarelis and
  Marmarelis}{1978}]{marmarelis78}
Marmarelis P, Marmarelis V (1978)
\newblock {\em Analysis of Physiological Systems: The White-Noise Approach}
\newblock Plenum, New York.

\bibitem[\protect\citeauthoryear{Mastronarde}{1983a}]{Mastronarde83I}
Mastronarde DN (1983a)
\newblock Correlated firing of cat retinal ganglion cells. {I}. {S}pontaneously
  active inputs to {X}- and {Y}-cells.
\newblock {\em J Neurophysiol}~49:\mbox{303--324}.

\bibitem[\protect\citeauthoryear{Mastronarde}{1983b}]{Mastronarde83II}
Mastronarde DN (1983b)
\newblock Correlated firing of cat retinal ganglion cells. {II}. {R}esponses of
  {X}- and {Y}-cells to single quantal events.
\newblock {\em J Neurophysiol}~49:\mbox{325--349}.

\bibitem[\protect\citeauthoryear{Milner}{1974}]{milner74}
Milner P (1974)
\newblock A model for visual shape recognition.
\newblock {\em Psychol Rev}~81:\mbox{521--535}.

\bibitem[\protect\citeauthoryear{Mitchell \bgroup et al.\egroup
  }{2009}]{mitchell09}
Mitchell JF, Sundberg KA, Reynolds JH (2009)
\newblock Spatial attention decorrelates intrinsic activity fluctuations in
  macaque area {V}4.
\newblock {\em Neuron}~63:\mbox{879--888}.

\bibitem[\protect\citeauthoryear{Nirenberg  \bgroup et al.\egroup
  }{2001}]{Nirenberg01}
Nirenberg S, Carcieri SM, Jacobs AL, Latham PE (2001)
\newblock Retinal ganglion cells act largely as independent encoders.
\newblock {\em Nature}~411:\mbox{689--701}.

\bibitem[\protect\citeauthoryear{Nirenberg and Latham}{2003}]{Nirenberg03}
Nirenberg S, Latham PE (2003)
\newblock Decoding neuronal spike trains: How important are correlations?
\newblock {\em Proc Natl Acad Sci USA}~100:\mbox{7348--7353}.

\bibitem[\protect\citeauthoryear{Ohiorhenuan  \bgroup et al.\egroup
  }{2010}]{Ohiorhenuan10}
Ohiorhenuan IE, Mechler F, Purpura KP, Schmid AM, Hu Q, Victor JD (2010)
\newblock Sparse coding and high-order correlations in fine-scale cortical
  networks.
\newblock {\em Nature}~466:\mbox{617--621}.

\bibitem[\protect\citeauthoryear{Oizumi  \bgroup et al.\egroup
  }{2010}]{oizumi10}
Oizumi M, Ishii T, Ishibashi K, Hosoya T, Okada M (2010)
\newblock Mismatched decoding in the brain.
\newblock {\em J Neurosci}~30:\mbox{4815--4826}.

\bibitem[\protect\citeauthoryear{Oizumi  \bgroup et al.\egroup
  }{2008}]{oizumi09}
Oizumi M, Ishii T, Ishibashi K, Hosoya T, Okada M (2008)
\newblock A general framework for investigating how far the decoding process in
  the brain can be simplified
\newblock In Koller D, Schuurmans D, Bengio Y, Bottou L, editors, {\em Advances
  in Neural Information Processing Systems 21}, \mbox{pp. 1225--1232}.

\bibitem[\protect\citeauthoryear{Optican and Richmond}{1987}]{optican87}
Optican LM, Richmond BJ (1987)
\newblock Temporal encoding of two-dimensional patterns by single units in
  primate inferior temporal cortex. {III}. information theoretic analysis.
\newblock {\em J Neurophysiol}~57:\mbox{162--178}.

\bibitem[\protect\citeauthoryear{Petersen \bgroup et al.\egroup
  }{2001}]{Petersen01}
Petersen RS, Panzeri S, Diamond ME (2001)
\newblock Population coding of stimulus location in rat somatosensory cortex.
\newblock {\em Neuron}~32:\mbox{503--514}.

\bibitem[\protect\citeauthoryear{Pillow  \bgroup et al.\egroup
  }{2008}]{Pillow08}
Pillow JW, Shlens J, Paninski L, Sher A, Litke AM, Chichilnisky EJ, Simoncelli
  EP (2008)
\newblock Spatio-temporal correlations and visual signalling in a complete
  neuronal population.
\newblock {\em Nature}~454:\mbox{995--999}.

\bibitem[\protect\citeauthoryear{Pouget \bgroup et al.\egroup
  }{2003}]{pouget03}
Pouget A, Dayan P, Zemel RS (2003)
\newblock Inference and computation with population codes.
\newblock {\em Annu Rev Neurosci}~26:\mbox{381--410}.

\bibitem[\protect\citeauthoryear{Richmond and Optican}{1990}]{richmond90}
Richmond BJ, Optican LM (1990)
\newblock Temporal encoding of two-dimensional patterns by single units in
  primate primary visual cortex. {II}. {I}nformation transmission.
\newblock {\em J Neurophysiol}~64:\mbox{370--380}.

\bibitem[\protect\citeauthoryear{Rodieck}{1967}]{rodieck67}
Rodieck R (1967)
\newblock Maintained activity of cat retinal ganglion cells.
\newblock {\em J Neurophysiol}~30:\mbox{1043--1071}.

\bibitem[\protect\citeauthoryear{Roudi \bgroup et al.\egroup }{2009a}]{Roudi09}
Roudi Y, Nirenberg S, Latham PE (2009a)
\newblock Pairwise maximum entropy models for studying large biological
  systems: when they can work and when they can't.
\newblock {\em PLoS Comput Biol}~5:\mbox{e1000380}.

\bibitem[\protect\citeauthoryear{Roudi \bgroup et al.\egroup
  }{2009b}]{Roudi09b}
Roudi Y, Tyrcha J, Hertz J (2009b)
\newblock Ising model for neural data: Model quality and approximate methods
  for extracting functional connectivity.
\newblock {\em Phys. Rev. E}~79:\mbox{051915}.

\bibitem[\protect\citeauthoryear{Roudi \bgroup et al.\egroup
  }{2009c}]{Roudi09c}
Roudi Y, Aurell E, Hertz JA (2009c)
\newblock Statistical physics of pairwise probability models.
\newblock {\em Front. Comput. Neurosci.}~3:\mbox{22}.

\bibitem[\protect\citeauthoryear{Schneidman  \bgroup et al.\egroup
  }{2006}]{Schneidman06}
Schneidman E, Berry M, Segev R, Bialek W (2006)
\newblock Weak pairwise correlations imply strongly correlated network states
  in a neural population.
\newblock {\em Nature}~440:\mbox{1007--1012}.

\bibitem[\protect\citeauthoryear{Schneidman \bgroup et al.\egroup
  }{2003a}]{schneidman03a}
Schneidman E, Bialek W, Berry MJ (2003a)
\newblock Synergy, redundancy, and independence in population codes.
\newblock {\em J Neurosci}~23:\mbox{11539--11553}.

\bibitem[\protect\citeauthoryear{Schneidman  \bgroup et al.\egroup
  }{2003b}]{schneidman03b}
Schneidman E, Still S, Berry MJ, Bialek W (2003b)
\newblock Network information and connected correlations.
\newblock {\em Phys Rev Lett}~91:\mbox{238701--238701}.

\bibitem[\protect\citeauthoryear{Shadlen  \bgroup et al.\egroup
  }{1996}]{Shadlen96}
Shadlen MN, Britten KH, Newsome WT, Movshon JA (1996)
\newblock A computational analysis of the relationship between neuronal and
  behavioral responses to visual motion.
\newblock {\em J Neurosci}~16:\mbox{1486--1510}.

\bibitem[\protect\citeauthoryear{Shannon and Weaver}{1949}]{shannon49}
Shannon C, Weaver W (1949)
\newblock {\em The mathematical theory of communication}
\newblock University of Illinois Press, Urbana, Illinois.

\bibitem[\protect\citeauthoryear{Shlens  \bgroup et al.\egroup
  }{2009}]{shlens09}
Shlens J, Field GD, Gauthier JL, Greschner M, Sher A, Litke AM, Chichilnisky EJ
  (2009)
\newblock The structure of large-scale synchronized firing in primate retina.
\newblock {\em J Neurosci}~29:\mbox{5022--5031}.

\bibitem[\protect\citeauthoryear{Shlens  \bgroup et al.\egroup
  }{2006}]{Shlens06}
Shlens J, Field G, Gauthier J, Grivich M, Petrusca D, Sher A, Litke A,
  Chichilnisky E (2006)
\newblock The structure of multi-neuron firing patterns in primate retina.
\newblock {\em J Neurosci}~26:\mbox{8254--8266}.

\bibitem[\protect\citeauthoryear{Sompolinsky  \bgroup et al.\egroup
  }{2001}]{Sompolinsky01}
Sompolinsky H, Yoon H, Kang K, Shamir M (2001)
\newblock Population coding in neuronal systems with correlated noise.
\newblock {\em Phys. Rev. E}~64:\mbox{051904}.

\bibitem[\protect\citeauthoryear{Staude \bgroup et al.\egroup
  }{2010}]{Staude10}
Staude B, Gr{\"u}n S, Rotter S (2010)
\newblock Higher-order correlations in non-stationary parallel spike trains:
  statistical modeling and inference.
\newblock {\em Front Comput Neurosci}~4.

\bibitem[\protect\citeauthoryear{Tang  \bgroup et al.\egroup }{2008}]{tang08}
Tang A, Jackson D, Hobbs J, Chen W, Smith JL, Patel H, Prieto A, Petrusca D,
  Grivich MI, Sher A, Hottowy P, Dabrowski W, Litke AM, Beggs JM (2008)
\newblock A maximum entropy model applied to spatial and temporal correlations
  from cortical networks in vitro.
\newblock {\em J Neurosci}~28:\mbox{505--518}.

\bibitem[\protect\citeauthoryear{Tka\v{c}ik}{2007}]{tkacik07}
Tka\v{c}ik G (2007)
\newblock Information flow in biological networks.
\newblock Ph.D. diss., Princeton University.

\bibitem[\protect\citeauthoryear{Tka\v{c}ik  \bgroup et al.\egroup
  }{2006}]{tkacik06}
Tka\v{c}ik G, Schneidman E, Berry M, Bialek W (2006)
\newblock Ising models for networks of real neurons.
\newblock {\em arXiv:q-bio/0611072v1}~.

\bibitem[\protect\citeauthoryear{Truccolo  \bgroup et al.\egroup
  }{2005}]{Truccolo05}
Truccolo W, Eden U, Fellows M, Donoghue J, Brown E (2005)
\newblock A point process framework for relating neural spiking activity to
  spiking history, neural ensemble and extrinsic covariate effects.
\newblock {\em J Neurophys}~93:\mbox{1074--1089}.

\bibitem[\protect\citeauthoryear{Victor}{2005}]{Victor05}
Victor JD (2005)
\newblock Spike train metrics.
\newblock {\em Current Opinion in Neurobiology}~15:\mbox{585--592}.

\bibitem[\protect\citeauthoryear{Victor and Purpura}{1996}]{victor96}
Victor JD, Purpura KP (1996)
\newblock Nature and precision of temporal coding in visual cortex: a
  metric-space analysis.
\newblock {\em J Neurophysiol}~76:\mbox{1310--1326}.

\bibitem[\protect\citeauthoryear{von~der Malsburg}{1981}]{vonderMalsburg81}
von~der Malsburg C (1981)
\newblock The correlation theory of brain function.
\newblock {\em MPI Biophysical Chemistry, Internal Report 81–2}~.

\bibitem[\protect\citeauthoryear{Wiener and Richmond}{1999}]{wiener99}
Wiener MC, Richmond BJ (1999)
\newblock Using response models to estimate channel capacity for neuronal
  classification of stationary visual stimuli using temporal coding.
\newblock {\em J Neurophysiol}~82:\mbox{2861--2875}.

\bibitem[\protect\citeauthoryear{Wu \bgroup et al.\egroup }{2001}]{wu01}
Wu S, Nakahara H, Amari S (2001)
\newblock Population coding with correlation and an unfaithful model.
\newblock {\em Neural Comput}~13:\mbox{775--797}.

\bibitem[\protect\citeauthoryear{Wu  \bgroup et al.\egroup }{2000}]{Wu00}
Wu S, Nakahara H, Murata N, Amari S (2000)
\newblock Population decoding based on an unfaithful model
\newblock In {\em Advances in neural information processing systems}, \mbox{pp.
  167--173}, Cambridge, MA: MIT press.

\bibitem[\protect\citeauthoryear{Yu  \bgroup et al.\egroup }{2008}]{yu08}
Yu S, Huang D, Singer W, Nikolic D (2008)
\newblock A small world of neuronal synchrony.
\newblock {\em Cereb Cortex}~18:\mbox{2891--2901}.

\bibitem[\protect\citeauthoryear{Zohary \bgroup et al.\egroup
  }{1994}]{Zohary94}
Zohary E, Shadlen MN, Newsome WT (1994)
\newblock Correlated neuronal discharge rate and its implications for
  psychophysical performance.
\newblock {\em Nature}~370:\mbox{140--143}.
\end{thebibliography}
\end{document}